\documentclass[12pt]{iopart}
\usepackage[utf8]{inputenc}
\usepackage{graphicx,color}
\graphicspath{{figuras/}}
\usepackage{subfigure}
\usepackage{epstopdf}
\usepackage{url}
\usepackage{cite}
\usepackage{iopams}   
\usepackage{caption}
\usepackage{soul}

\begin{document}
\title{The Moon, a disk or a sphere?}

\author{E.Seperuelo Duarte$^{1}$, A.T.Mota$^{2}$, J.R. de Carvalho$^{2}$, R.C. Xavier$^{2}$ and P.V.S.Souza$^{2}$}

\address{$^1$Instituto Federal de Educação, Ciência e Tecnologia do Estado do Rio de Janeiro, Campus Nilópolis, 26530-060 Nilópolis, Rio de Janeiro, Brazil}

\address{$^2$Instituto Federal de Educação, Ciência e Tecnologia do Estado do Rio de Janeiro, Campus Volta Redonda, 27215-350 Volta Redonda, Rio de Janeiro, Brazil}

\ead{paulo.victor@ifrj.edu.br}
\vspace{10pt}
\begin{indented}
\item[]June 2021
\end{indented}

\begin{abstract}
	{In this paper, we present a physical modeling activity whose objective is to allow students to determine the differences between a disk and a sphere using pure scientific criteria. Thereunto, we reproduce the Sun-Earth-Moon system with low-cost materials and compare the illumination effects on the Moon considering two possible shapes for it (a sphere and a disk). {The analysis is based on the shape of the {terminator line} produced in each case as a function of the illumination angle. The results obtained are first discussed and then applied so that one can interpret the observed patterns in the illumination effects of other celestial bodies, such as Venus or even the Earth. Thereby, the activity can be very useful to unmask the unscientific idea of Flat Earth. The entire activity is easily replicable and it may be useful to promote a more realistic view of science and its methods. }}
\end{abstract}
\noindent \textit{ Keywords}: {Scientific modeling, Sun-Earth-Moon system, Flat Earth, Terminator line;}

\maketitle
%
\section{Introduction}
Human knowledge has developed over time with significant contributions from the sciences. They promoted the advancement of technologies that today can make our lives more comfortable and safer. Astronomy, known as the oldest of sciences, played a decisive role in this evolution and the research carried out in this area will certainly continue to contribute to society \cite{pannekoek1989history}. 

In particular, mankind's interest in the Moon is old. The periodic lunation event, for example, is linked to what we define today as a month. In addition, the Moon has always been an object of admiration and study. The search for understanding how it originated tells our own story. From the arrival of man at the Moon in 1969 to the current Space exploration, Astronomy has contributed to the advancement of technologies and basic science \cite{parker1950calendars,owano2013scotland,sagdeev2008united,schefter2010race}. 

However, a new anti-scientific wave has called into question the results obtained by science on several topics. One of these themes is related to the sphericity of the Earth. Flat Earth movement is being defended once again by people who are unaware of scientific procedures and insist on defending opinions without scientific evidence \cite{mcintyre2015respecting,landrum_olshansky_2019,mcintyre2019calling,Brazil_2020,hotez2021anti}.

In this context, the role of science education and, specifically, astronomy teaching, is fundamental to promote the acknowledgment of scientific research methods and their use as a tool to understand the current world \cite{marschall1996bringing,ellery2012measuring, carroll2013using,tort2013exercise,noordeh2014simulating,BZOV_2018,Kuzii_2019,francca2019propostas,treff2019image,caerols2020estimating}.
Accordingly, we present a physical modeling activity that allows students to determine the differences between a disk and a sphere using pure scientific criteria. In particular, we reproduce the Sun-Earth-Moon system and compare the illumination effects on the Moon considering two shapes: (i) a sphere and (ii) a disk. The terminator line, or twilight zone, is a moving line that divides the daylight side and the dark night side of a planetary body. The lunar terminator is the division between the illuminated and dark hemispheres of the Moon \cite{liu2009solar,de2018angle}, as shown in Fig. 1. 

 \begin{figure}[h!]
    \centering
    \includegraphics[scale=0.26]{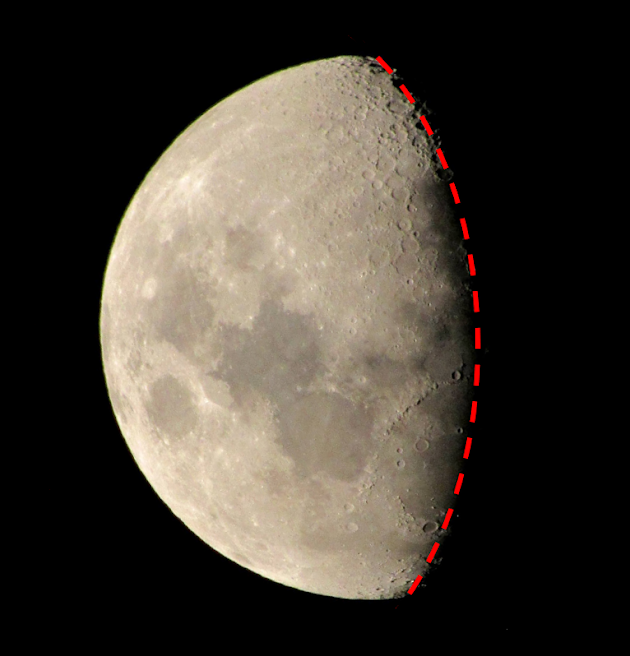}
    \caption{{This is a picture of the Moon taken from Earth.} The dashed line indicates the Lunar terminator line}
  \label{1} 
  \end{figure}
  
In the activity, the relative positions of the Sun, the Earth, and the Moon {in the model} were varied in a way to simulate the phases of the Moon. The aim was to verify the shape changes of the terminator line and associate it with the Moon shape. The lunar phases were registered by a cell phone camera and the images can be used in the classroom. The comparison with the effects obtained when the disk was used to represent the Moon's surface makes it clear that the best shape to represent the lunar surface is a sphere. The materials needed to perform the experiment have a very low cost so that it can be reproduced in any environment with the conditions described below.

\section{The experimental procedure}

In this section, we present the performed experiment. Fig. 2(A) shows the experimental setup.

 \begin{figure}[h!]
    \centering
    \includegraphics[scale=0.28]{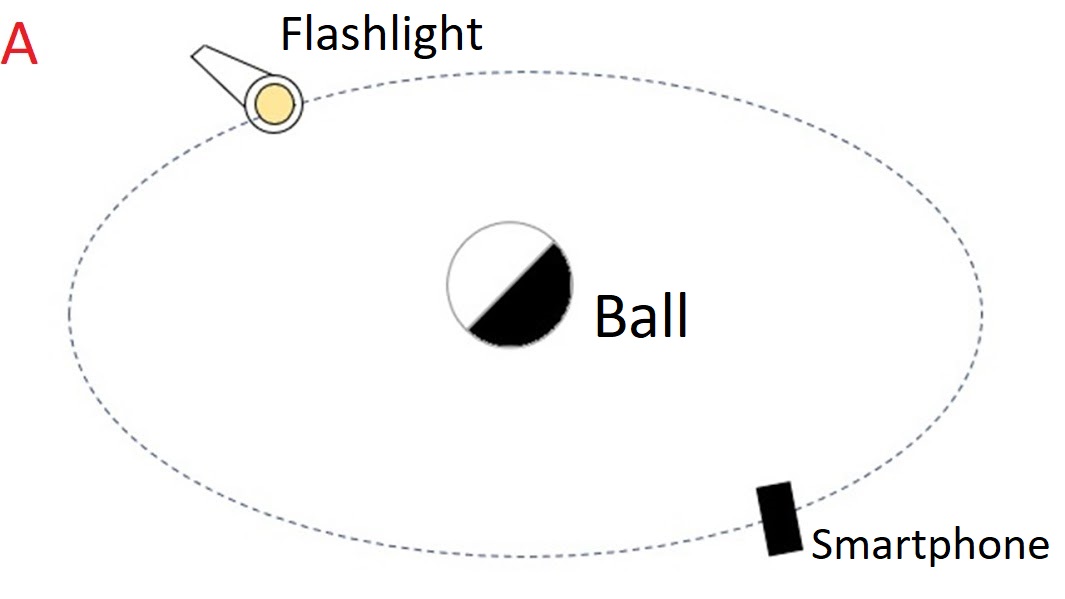}
    \hspace{-10pt}
         \includegraphics[scale=0.31]{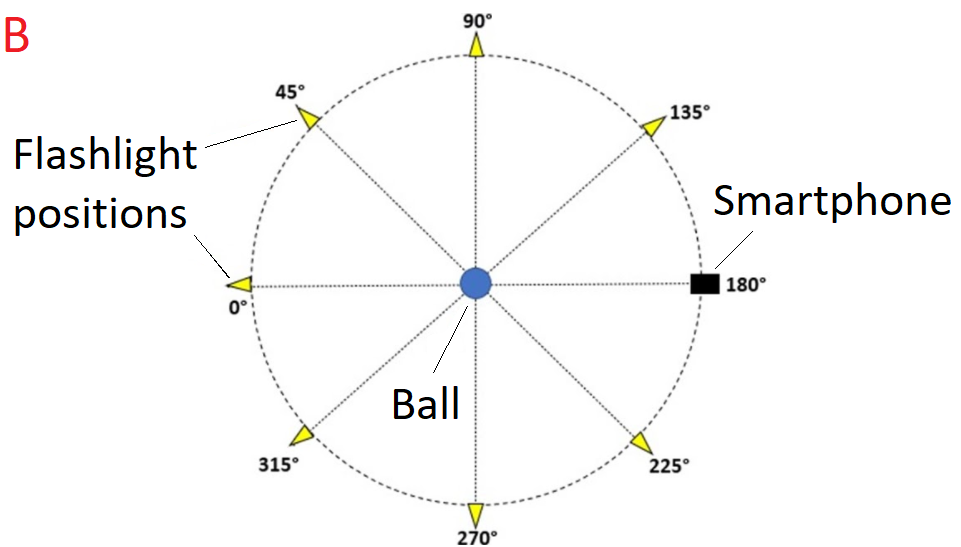}
    \caption{(A) The experimental setup; (B) The different positions considered for the flashlight to reproduce the relative positions between the Sun, the Earth and the Moon. {The ball} in the figures are the sphere and the disk.}
  \label{2} 
  \end{figure}

The materials used in this experiment are:
\begin{itemize}
    \item Styrofoam ball (60 mm);
    \item Flashlight with 2.5 cm of diameter;
    \item Cell phone holder;
    \item Sewing thread to hang the sphere;
    \item Styrofoam cardboard
    \item Dark fabrics to darken the room;
    \item Measuring tape;
    \item Samsung phone camera settings: Camera: Samsung SM - J701MT; Aperture: F 1.9; Focal length: 3.70 mm; ISO: 800; Exposure time: 1/10 s; White balance: automatic.
\end{itemize}

Choosing the flashlight depends on the size of the ball. Once we used a ball with a diameter of 60 mm, it was necessary to select a flashlight that illuminates the entire surface of the sphere with collimated light, so that it was easily possible to reproduce all phases of the {Moon}. In this case, the flashlight lamp used was a CREE LED High Power (58000W).   

The cell phone stood in front of the ball. The distance between the ball's center and the camera was 73cm. Hence, the light was coming from the flashlight that was being moved around it to get as close as possible to reproduce the phases of the Moon. The closer the light was to the ball, the better terminator line visualization was obtained. In order to avoid diffuse reflection, the room´s walls were covered by a dark fabric. The ``Moon" object was suspended by a sewing thread so its height to the floor is the same as the cellphone camera.  



Then, we took 8 photos of the sphere, one for each angle of incidence: 0º, 45º, 90º, 135º, 180º, 225º, 270º, and 315º, as shown in Fig. 2(B). We repeated the procedure replacing the sphere for a disk {which remains fixed and whose surface is always perpendicular to the direction defined by the straight line connecting the disc and the camera.} In the next section, the results obtained are compared and interpreted.

\section{The results}

The cycle of the Moon has four main phases: new Moon, first quarter, full Moon, and third quarter. There are intermediate phases in the intervals between the main phases: waxing crescent (new Moon to the first quarter), waxing gibbous (first quarter to full Moon), waning gibbous (full Moon to the third quarter), and waning crescent (third quarter to new Moon). {Henceforth, the model components will be referred with lower case and the real objects with capital letters.} Fig. 3 shows the sequence of images obtained, from the new {moon} to the waning crescent {moon}. The images were labeled by the letter S (sphere) and by letter D (disk) followed by the $\theta$ angle value. Note that a terminator line can be produced only when a sphere is used.

 \begin{figure}[h!]
    
    \includegraphics[scale=0.22]{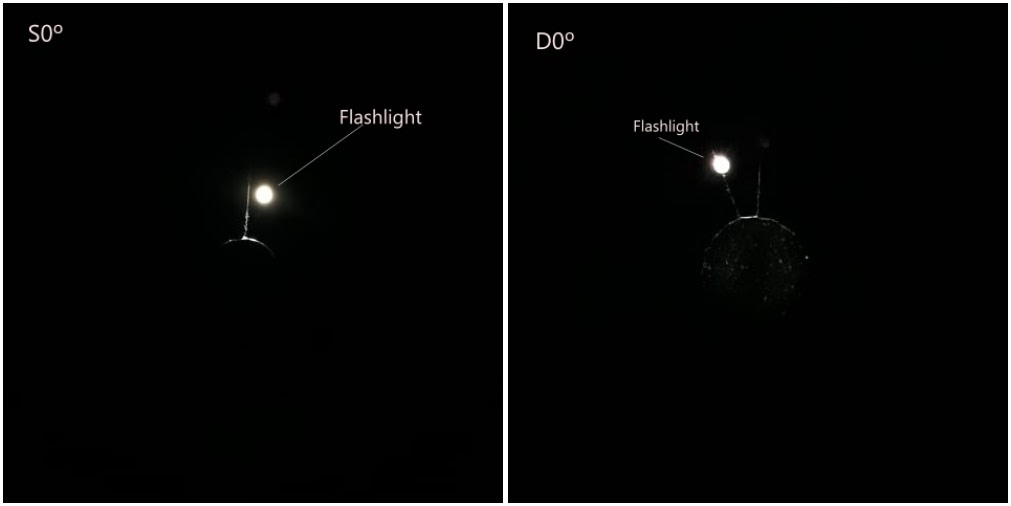} 
    \hspace{-10pt}
    \includegraphics[scale=0.22]{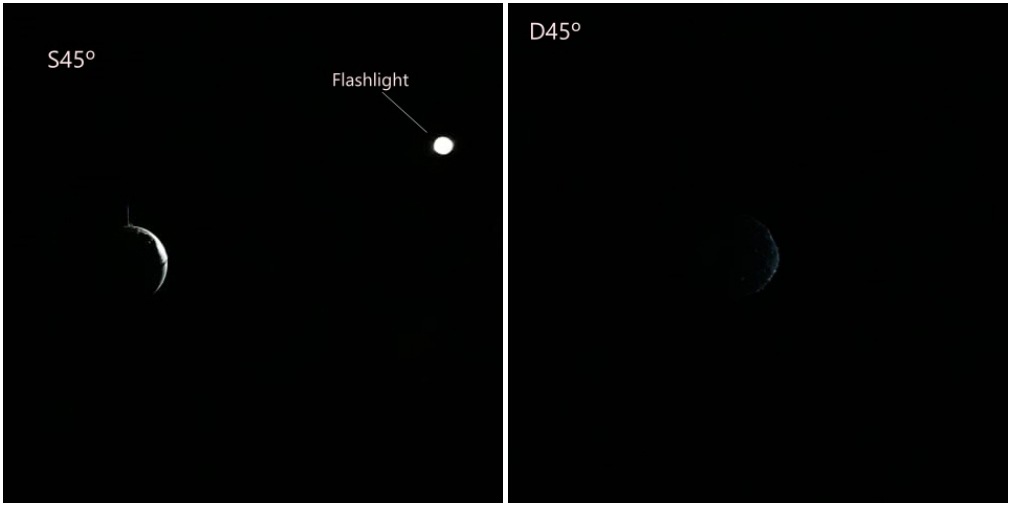} 

         \includegraphics[scale=0.22]{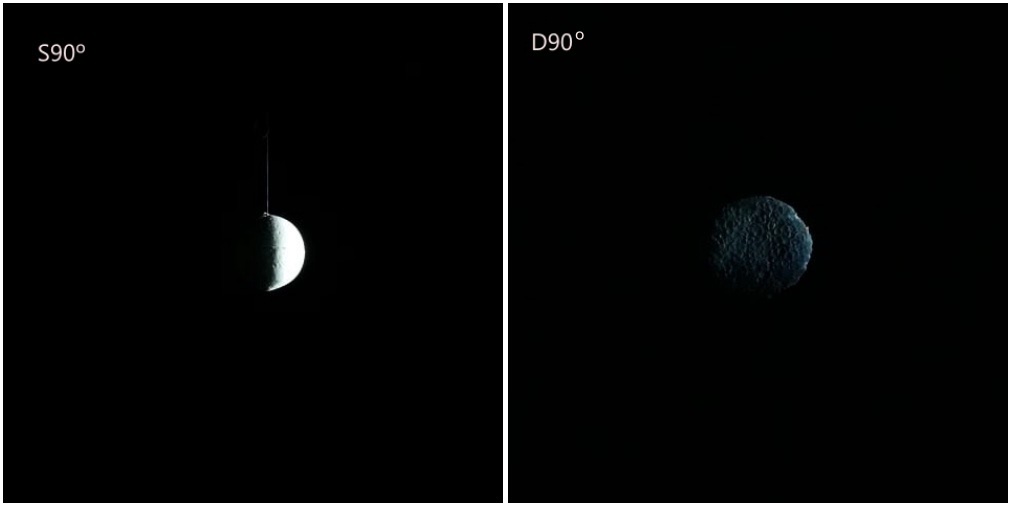} 
\hspace{-10pt}
     \includegraphics[scale=0.22]{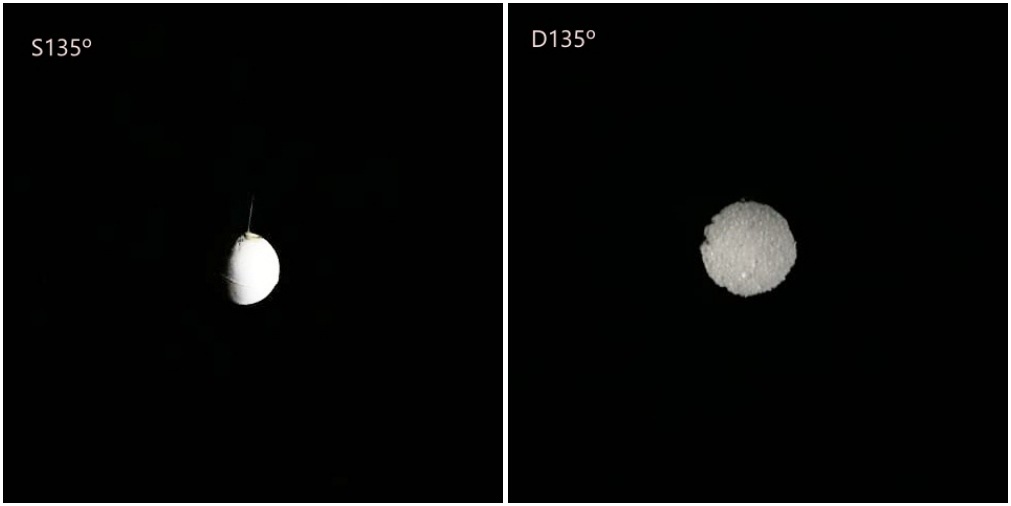} 
     
     \includegraphics[scale=0.22]{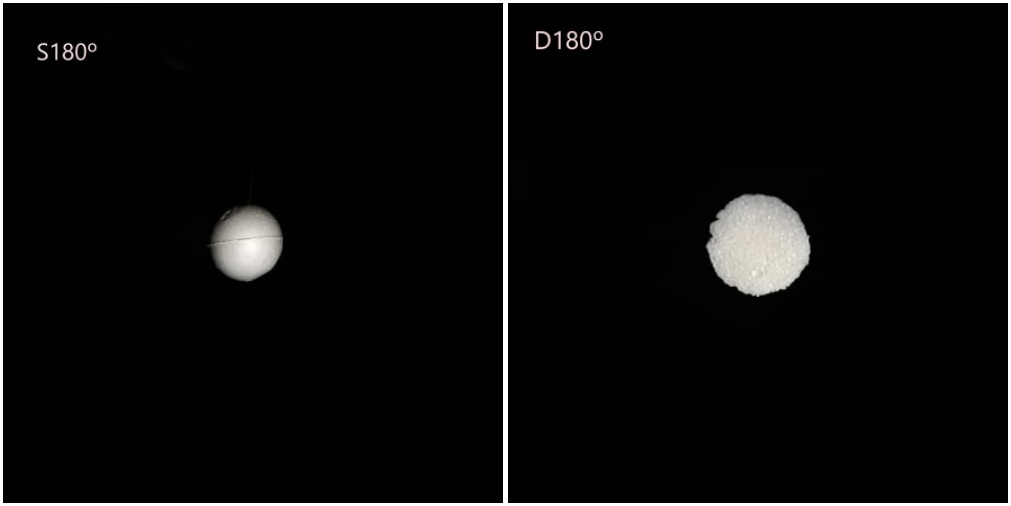} 
    \hspace{-10pt}
     \includegraphics[scale=0.22]{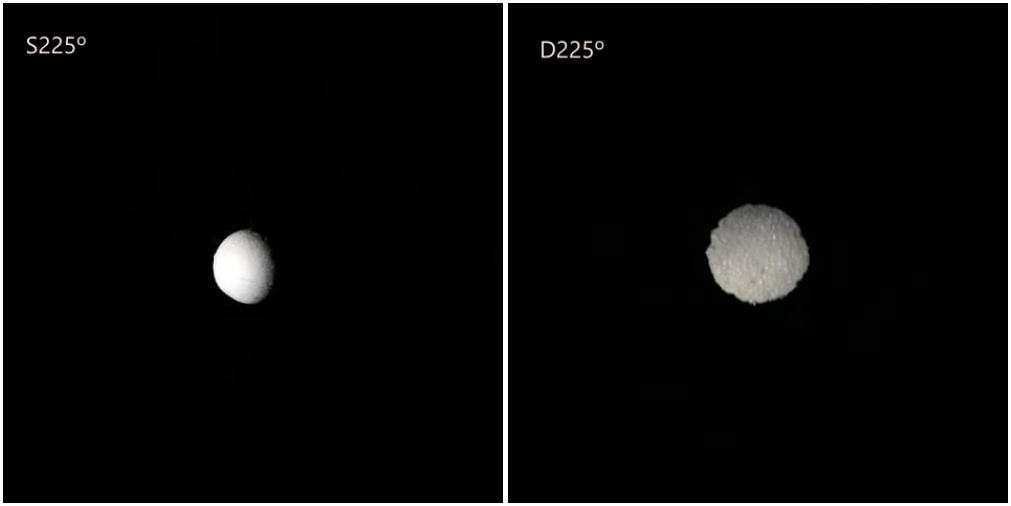} 
     
          \includegraphics[scale=0.22]{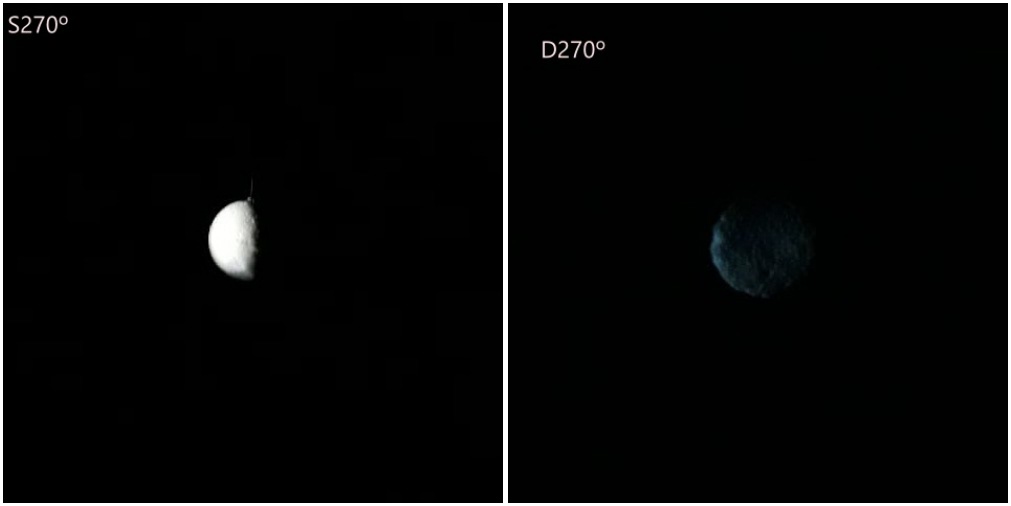} 
     \hspace{-10pt}
     \includegraphics[scale=0.22]{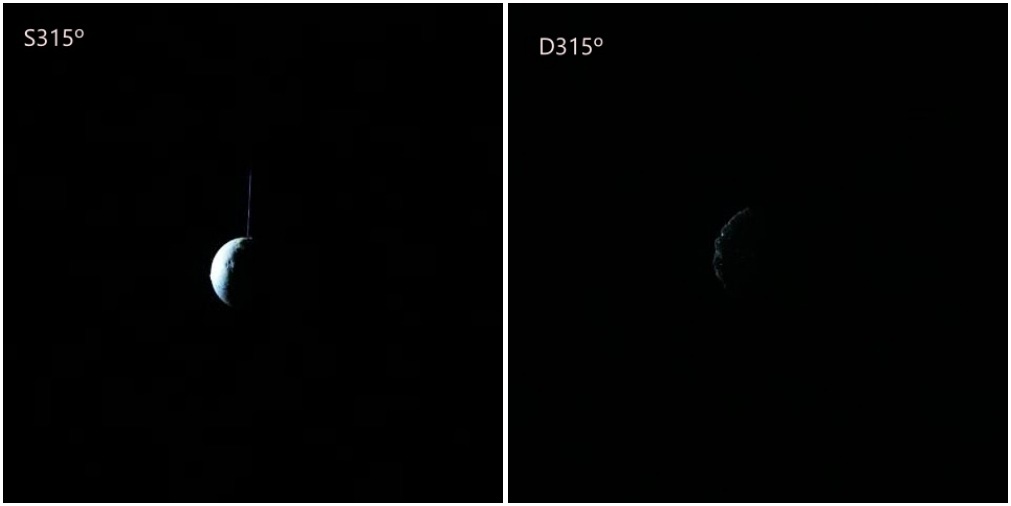} 

    \caption{{Pictures obtained for the sphere and the disk as a function of some angles of incidence of light, the same as those shown in figure 2B. In each photo, the letters S(D) correspond to the photos obtained for the sphere (disk). The angle of incidence of the light is indicated by the number right next to each letter S or D.}}
  \label{3} 
  \end{figure}

\subsection{The spherical {moon}}
The results obtained when the Moon is represented by a sphere is shown in Fig. 3 (S-panels). At the new {moon}, the {sun}, the {moon}, and the {earth} are aligned in a conjunction ($\theta$ = 0º) and the {moon's} visible side is not being illuminated by the sun. The spotlight seen in this image (S0º) is produced by the flashlight used in the experiment. 

During the waxing crescent phase ($\theta$ = 45º), the {moon} is being illuminated by the sun from its {right visible side as seen from the earth} and the terminator line has a concave shape. The concavity degree decreases as the phase evolves and the terminator line becomes a straight line in the first quarter.

After the first quarter ($\theta$ = 90º), the {left} visible side  of the {moon} starts to be illuminated. The terminator of the waxing gibbous {moon} ($\theta$ = 135º) is a convex line with the concavity degree increasing as it moves toward the {moon´s left} border. In the full {moon} ($\theta$ = 180º), the visible side is totally illuminated by the sun and the terminator is localized in the circumferential border of the {moon}.

When the waning gibbous ($\theta$ = 225º) phase starts, the terminator line rises from the {right} border with a convex shape. It moves toward the center of the {moon's} visible side decreasing its concavity. In the third quarter ($\theta$ = 270º), the terminator line becomes straight again. At this moment, the sun is illuminating only the {moon´s left} visible side.
Finally, in the last phase quarter, the waning crescent ($\theta$ = 315º), the terminator is a concave line again moving toward the {moon´s left} border and increasing its concavity.

\subsection{The Flat {moon}}

The results obtained when the Moon is represented by a disk are shown in Fig. 3 (D-panels). Note that for the flashlight at the angles of 0º, 45º and 315º, the visible side of the {moon} is not being illuminated and no terminator line is produced. At angles of 90º and 270º, the flashlight is tangent to the disk surface and some diffuse reflection is observed due to inhomogeneities on it. The disk is illuminated only when the flashlight is at angles of 135º, 180º, and 225º.

In none of the cases, a terminator line is produced on the disk surface. The lunar phases are reduced to two: a new {moon} and a full {moon}. However, a change in brightness for different angles, with the maximum brightness at the full {moon}, is observed. It is not possible to reproduce the whole cycle of lunar phases with this model, that is, using a disk to represent the Moon. In other words, it is impossible to produce a terminator line on the lunar surface with this setup, {that is, with a fixed disk illuminated at different angles.} 

\section{Illumination effects on other celestial bodies}

\begin{figure}[h!]
    \centering
    \includegraphics[scale=0.645]{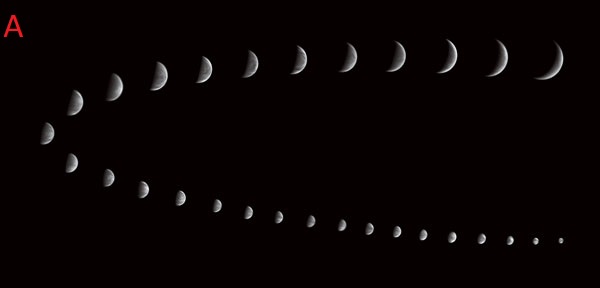}
    \includegraphics[scale=0.58]{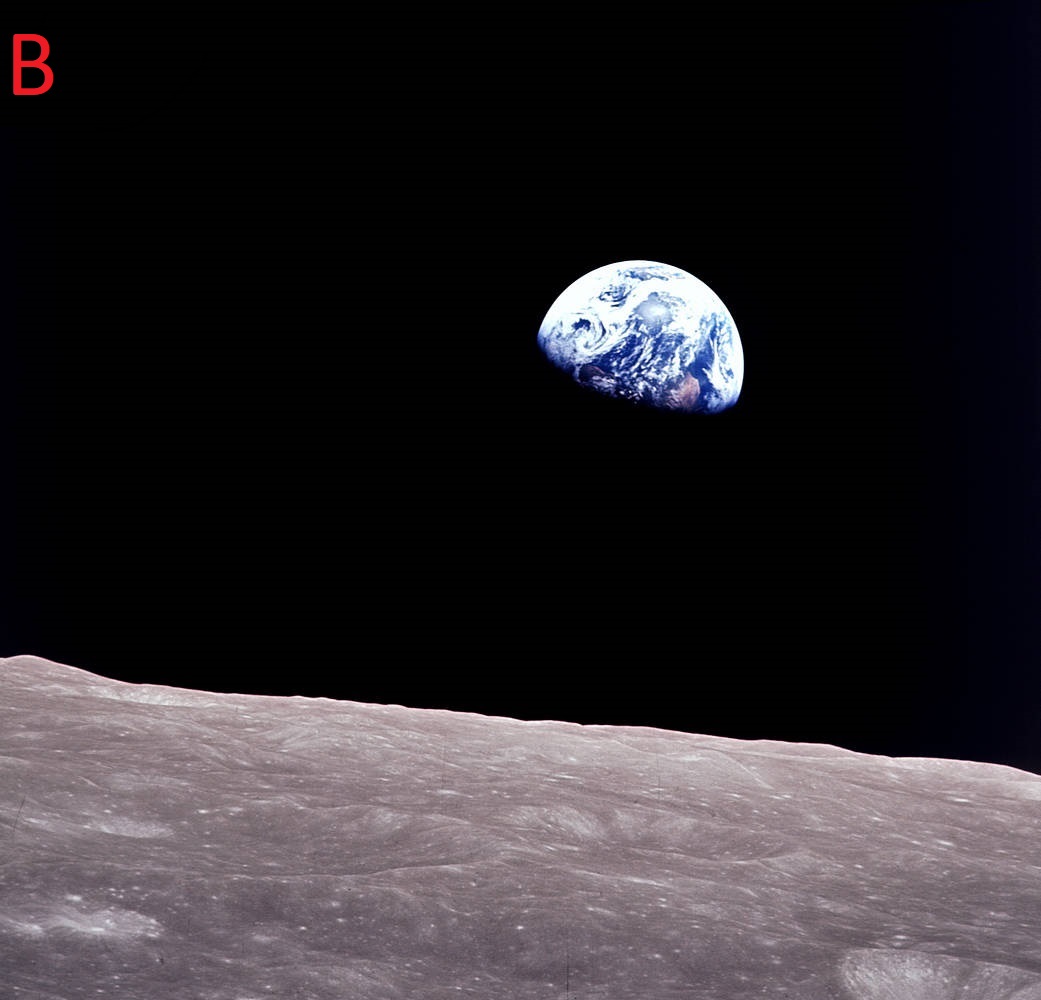}
    \caption{ (A) The planet Venus goes through a cycle of phases similar to the Moon. This sequence was recorded in 2007 by Sean Walker. Sky \& Telescope 2020. (B) The Earthrise taken by the astronaut Willian Anders aboard the Apollo 8 as the spacecraft circumnavigated the Moon in 1968. NASA.}
  \label{4} 
  \end{figure}

{The results obtained with the model may be extended to the investigation of the illumination effects on other celestial bodies, something that could be very interesting in the classroom, especially considering the recent resurgence of the Flat Earth movement.} Due to the great distances, the observation of the Solar System planets is a little more difficult than the observation of the Moon. However, it is possible to observe some planet features even though using small telescopes. Specifically, taking into account just the Solar System planets, the terminator line of Venus is the easiest to be observed (Figure \ref{4}a). Due to its internal orbit and its proximity, the relative positions of Venus, the Sun, and the Earth produce phases on the planet equivalent to the lunar phases.

  In the case of the Earth, only after the dawn of space exploration, it was possible to obtain images {of} the planet. The first color photograph of the Earth taken from the orbit of another celestial body (the Moon) was during the Apollo 8 mission on December 24th, 1968 by the astronaut Willian Anders (Figure \ref{4}b).

Clearly, a convex terminator line can be seen in this picture and, as the Moon moves in its orbit, we would observe a similar cycle of phases on the Earth. As discussed above, a curved surface model can reproduce this terminator line and its variations for the Earth phases.
Furthermore, based on the activity presented in this work and considering a spherical surface model, the Earth phase can be determined at the moment this picture was taken. 
Given that the Earth terminator line is a convex line crossing Africa and the North Hemisphere is located at the right, the Earthside that is being illuminated by the Sun is the west side\footnote{{This information could be confirmed, for instance, on \textcolor{blue}{\url{https://www.theguardian.com/artanddesign/2018/dec/22/behold-blue-plant-photograph-earthrise}}, accessed on June 8, 2021.}}. Since the Moon moves from west to east in its orbit, the Earth terminator line is moving up west. Thus, the Moon is moving toward the dark night side and the Earth phase in this picture would be waning gibbous Earth. 
The relative positions between the Sun, the Earth, and the Moon can be determined considering the angles defined in Fig. 2(B). A waning gibbous Earth is an intermediate phase between the full Earth and the third quarter. At full Earth, the Moon is between the Sun and the Earth {(Figure 2B with the sun at 0 degree)}. In the third quarter, the Sun and the Earth are 90 degrees apart in the lunar sky {(Figure 2B with the sun at 90 degrees)}. Thus, at waning gibbous Earth, the angle between the Sun and the Earth at the lunar sky is about {135 degrees (or 315 degrees anticlockwise) (Figure 2B with the sun at position 45 degrees)}. Then, at the moment the picture at Figure \ref{4}B was taken, the lunar phase observed on Earth was a waxing crescent Moon (Figure \ref{3}; S45º). {We would like to point out that this result can be confirmed by consulting a lunar calendar.}

\section{Final remarks}
In this work, we have presented a simple and easy to replicate physical modeling activity that allows students to determine the differences between a disk and a sphere. It incorporates several elements typical of the scientific method, such as observation, construction, and testing of hypotheses, modeling, forecasting, and establishment of conclusions based on critical thinking. Once the activity could provide students with an experience similar to that of scientists in their research laboratories, it may promote the construction of a more real vision about science as a tool that could support the decision-making process and a rational point of view facing the problems that permeate the twenty-one-century society.


\section{Acknowledgments}
{We would like to thank the support from Coordenação de Extensão - IFRJ Campus Nilópolis (grant agreement Edital 11/2019) and the first referee for the suggestions that certainly became the text clearer and pleasant.}

\section{References}

\bibliographystyle{unsrt} 
\bibliography{ref}

\begin{thebibliography}{10}

\bibitem{pannekoek1989history}
A~Pannekoek.
\newblock {\em A history of astronomy}.
\newblock Dover Publicatios, INC, New York, 1989.

\bibitem{parker1950calendars}
R~A Parker.
\newblock {\em The calendars of ancient Egypt}.
\newblock University of Chicago Press, Chicago, USA, 1950.

\bibitem{owano2013scotland}
N~Owano.
\newblock Scotland lunar-calendar find sparks stone age rethink.
\newblock {\em Phys. Org.}, 27, 2013.

\bibitem{sagdeev2008united}
R~Sagdeev, S~Eisenhower, and J~Lodgson.
\newblock United states-soviet space cooperation during the cold war.
\newblock {\em National Aeronautics and Space Administration}, 2008.
\newblock Accessed: 2021-02-26.

\bibitem{schefter2010race}
J~Schefter.
\newblock {\em The race: the complete true story of how America beat Russia to
  the Moon}.
\newblock Anchor, 2010.

\bibitem{mcintyre2015respecting}
L~McIntyre.
\newblock {\em Respecting truth: Willful ignorance in the internet age}.
\newblock Routledge, 2015.

\bibitem{landrum_olshansky_2019}
A~R Landrum and A~Olshansky.
\newblock The role of conspiracy mentality in denial of science and
  susceptibility to viral deception about science.
\newblock {\em Politics and the Life Sciences}, 38(2):193–209, 2019.

\bibitem{mcintyre2019calling}
L~McIntyre.
\newblock Calling all physicist.
\newblock {\em American Journal of Physics}, 87(9):694, 2019.

\bibitem{Brazil_2020}
R~Brazil.
\newblock Fighting flat-earth theory.
\newblock {\em Physics World}, 33(7):35--39, jul 2020.

\bibitem{hotez2021anti}
P~J Hotez.
\newblock Anti-science kills: from soviet embrace of pseudoscience to
  accelerated attacks on us biomedicine.
\newblock {\em PLoS biology}, 19(1):e3001068, 2021.

\bibitem{marschall1996bringing}
L~A Marschall.
\newblock Bringing the moon into the classroom.
\newblock {\em Physics Teacher}, 34(6):360--61, 1996.

\bibitem{ellery2012measuring}
A~Ellery and S~Hughes.
\newblock Measuring the apparent size of the moon with a digital camera.
\newblock {\em Physics Education}, 47(5):616, 2012.

\bibitem{carroll2013using}
J~Carroll and S~Hughes.
\newblock Using a video camera to measure the radius of the earth.
\newblock {\em Physics Education}, 48(6):731, 2013.

\bibitem{tort2013exercise}
A~C Tort.
\newblock An exercise on gauss' law for gravitation: The flat earth model.
\newblock {\em arXiv preprint arXiv:1305.0393}, 2013.

\bibitem{noordeh2014simulating}
E~Noordeh, P~Hall, and M~Cuk.
\newblock Simulating the phases of the moon shortly after its formation.
\newblock {\em The Physics Teacher}, 52(4):239--240, 2014.

\bibitem{BZOV_2018}
L~B{\v{r}}{\'{\i}}zov{\'{a}}, K~Gerbec, J~{\v{S}}auer, and J~{\v{S}}l{\'{e}}gr.
\newblock Flat earth theory: an exercise in critical thinking.
\newblock {\em Physics Education}, 53(4):045014, may 2018.

\bibitem{Kuzii_2019}
O~Kuzii and A~Rovenchak.
\newblock What the gravitation of a flat earth would look like and why thus the
  earth is not actually flat.
\newblock {\em European Journal of Physics}, 40(3):035008, apr 2019.

\bibitem{francca2019propostas}
M~J~M Fran{\c{c}}a.
\newblock Propostas de atividades de simula{\c{c}}{\~a}o e videoan{\'a}lise de
  um brinquedo popular e uma atividade em astronomia.
\newblock Master's thesis, Universidade Federal do Rio de Janeiro. Rio de
  Janeiro, 2019.

\bibitem{treff2019image}
V~Treff, A~C Bertuola, and V~S Filho.
\newblock An image analysis method for calculating the moon’s orbital
  eccentricity.
\newblock {\em The Physics Teacher}, 57(8):562--564, 2019.

\bibitem{caerols2020estimating}
H~Caerols and F~A Asenjo.
\newblock Estimating the moon-to-earth radius ratio with a smartphone, a
  telescope, and an eclipse.
\newblock {\em The Physics Teacher}, 58(7):497--501, 2020.

\bibitem{liu2009solar}
H~Liu, H~L{\"u}hr, and S~Watanabe.
\newblock A solar terminator wave in thermospheric wind and density
  simultaneously observed by champ.
\newblock {\em Geophysical Research Letters}, 36(10), 2009.

\bibitem{de2018angle}
P~A de~Eulate~Mart{\'\i}n, D~C Gloria, and M~T Francisco.
\newblock Angle between terminator and meridian: flat geometry versus formulae
  of solar azimuth and an easy approach to the daylight map.
\newblock {\em European Journal of Physics}, 39(4):045805, 2018.

\end{thebibliography}

\end{document}